\documentstyle[aps,multicol,amssymb,psfig,float,times]{revtex}

\newcommand{\be}{\begin{equation}}
\newcommand{\ee}{\end{equation}}
\newcommand{\bea}{\begin{eqnarray}}
\newcommand{\eea}{\end{eqnarray}}

\def\p1{\pi_1}

\def\l{\lambda}
\def\f{\phi}

\def\r{\rho}

\def\pmas{\partial_+}
\def\pmen{\partial_-}

\begin{document}

\draft

%\preprint{
%\begin{tabular}{r} FTUV/99-72 \\ IFIC/99-75 \\ UB-ECM-PF 99/19
%\end{tabular}
%}

\title{Exact late time Hawking radiation and the
information loss problem\\
for evaporating near-extremal black holes}

\author{Alessandro Fabbri$^{\ast}$}
\address{
Dipartimento di Fisica dell'Universit\`a di Bologna and INFN
sezione di Bologna,\\
Via Irnerio 46, 40126 Bologna, Italy. }
\author{Diego J. Navarro$^{\dagger}$ and Jos\'e Navarro-Salas$^{\ddagger}$}
\address{Departamento de F\'{\i}sica Te\'orica and IFIC, Centro Mixto
Universidad de Valencia-CSIC.\\
Facultad de F\'{\i}sica, Universidad de Valencia, Burjassot-46100, Valencia,
Spain.}

\maketitle

\begin{abstract}

In this paper we investigate the effects of gravitational
backreaction for the late time Hawking radiation of evaporating
near-extremal black holes. This problem can be studied within the
framework of an effective one-loop solvable model on AdS$_2$. We
find that the Hawking flux goes down exponentially and it is
proportional to a parameter which depends on details of the
collapsing matter. This result seems to suggest that the
information of the initial state is not lost and that the boundary
of AdS$_2$ acts, at least at late times, as a  sort of stretched
horizon in the Reissner-Nordstr\"om spacetime.

\end{abstract}


\pacs{\tt FTUV-00-1204 IFIC/00-76
     hep-th/0012017}

\begin{multicols}{2}

\narrowtext

The discovery of black hole radiation \cite{h} has led to a long standing
debate concerning the suggestion \cite{h2} that the evaporation process
implies a loss of quantum coherence. This conclusion seems inevitable if one
assumes the propagation of quantum fields on a fixed classical background.
However, backreaction effects could change this picture. 't Hooft \cite{th}
suggested that for an asymptotic observer the interaction between the
infalling matter and the outgoing radiation could preserve the information
of the initial quantum state of the collapsing matter through non-local
effects. Within this alternative viewpoint it has also been proposed a
principle of complementarity \cite{th,S,STU,s2}, which states that the
simultaneous measurements made by an external observer and those made by an
infalling observer crossing the horizon are forbidden.\\

In this letter we shall analyze the evaporation process of a
Reissner-Nordstr\"om black hole near extremality in a way which is loosely
connected with  the principle of complementarity. According to it we cannot
have a detailed description of the physics near the horizon and,
simultaneously, far away from the black hole. We shall restrict the
Einstein-Maxwell theory in a region very close to the horizon. If the
physical configurations to be considered preserve the spherical symmetry
and are close to extremality the resulting effective theory turns out to be
equivalent to a solvable two-dimensional model. The effective model remains
solvable also at the one-loop quantum level and it has been studied in
\cite{F,F2}, where we found it natural to describe the evaporation of the
black hole from the point of view of an infalling observer very close to the
horizon. In this paper we shall improve our analysis and consider the same
process as it is seen by an asymptotic observer at late retarded times (this
is the part of future null infinity which can still be described by our 2d
model). The solution we will get is very different in form from the original
one (they indeed describe two very different regions of the spacetime),
however we will crucially impose that they naturally  match at one point,
i.e. at the end-point of the
evaporation where the solution becomes extremal. In contrast with
the standard picture, the Hawking flux goes down at late times and
it is not proportional to the total mass of the collapsing matter.
Instead, we find that it is proportional to a parameter which
admits an infinite series expansion in Planck constant and depends
on all the higher order momenta of the classical stress-tensor of
the incoming matter. At leading order this parameter is the total
mass. All this seems to suggest that the information of the
initial state will be then released out to future null infinity
during the evaporation process.\\

We start our analysis presenting the two-dimensional effective
theory that describes the near-horizon region of the
Einstein-Maxwell theory around extremality (the mass of the
extremal hole is $m_0\simeq ql^{-1}$, where $l^2=G$ is Newton's
constant). We refer to \cite{F,F2} (and references therein) for
the details and for a full description of the methods used in this
work . The classical action is given by the Jackiw-Teitelboim
model \cite{JT}
\be
\label{jtaction} I = \int d^2x \sqrt{-g} \left[ (R +
\frac{4}{l^2q^3}) \tilde{\f} - \frac{1}{2} |\nabla f|^2 \right] \, ,
\ee
where the two-dimensional fields $g_{ab}^{(2)}$ and $\tilde
\phi$ appearing in (\ref{jtaction}) are related to the
four-dimensional metric by the expression
\be
\label{4ds}
ds_{(4)}^2= \frac{2l}{r_0}ds_{(2)}^2 + (r_0^2 + 4l^2\tilde{\f}) d{\Omega}^2
\, ,
\ee
and $r_0 = lq$ is the extremal radius. The field $f$ represents a spherically
symmetric scalar field which propagates freely in the region very close to the
horizon.\\

To properly account for backreaction effects we have to consider the
corresponding one-loop effective theory. Therefore we have to correct
(\ref{jtaction}) by adding the Polyakov-Liouville term \cite{P}
\bea
\label{paction}
I &=& \int d^2x \sqrt{-g} \left(R \tilde{\phi} + 4
\lambda^2 \tilde{\phi} -\frac{1}{2} \sum_{i=1}^N |\nabla f_i|^2\right)
\nonumber \\
&-&  \frac{N\hbar}{96\pi} \int d^2x \sqrt{-g} R \; \square^{-1} R +
\frac{N\hbar}{12\pi} \int d^2x \sqrt{-g} \lambda^2 \, ,
\eea
where we have considered the presence of $N$ scalar fields to have a
well-defined theory in the large $N$ limit. Note that the Polyakov-Liouville
action has a cosmological constant term which has been fixed ($\lambda^2 =
l^{-2}q^{-3}$) to ensure that the extremal configuration remains a solution of
the quantum theory. In conformal gauge $ds^2=-e^{2\r}dx^+dx^-$ the equations
of motion derived from (\ref{paction}) are
\bea
\label{eq1}
2\pmas \pmen \r + \l^2 e^{2\r} &=& 0 \, , \\
\label{eq2}
\pmas \pmen \tilde{\f} + \l^2 \tilde{\f} e^{2\r} &=& 0 \, , \\
\label{eq3}
\pmas \pmen f_i &=& 0 \, , \\
\label{eq4}
-2\partial^2_{\pm} \tilde{\f} + 4 \partial_{\pm} \rho \partial_{\pm}
\tilde{\f} &=& T^f_{\pm \pm} - \frac{N\hbar}{12\pi} t_{\pm} - \\
&& \frac{N\hbar}{12\pi} \left( (\partial_{\pm} \r )^2 -
\partial_{\pm}^2 \r \right) \, , \nonumber
\eea
where the chiral functions $t_{\pm}(x^{\pm})$, coming from the non-locality of
the Polyakov-Liouville action, are related with the boundary conditions of
the theory associated with the corresponding observers. The equation
(\ref{eq1}) is the Liouville equation with a negative cosmological constant.
It has a unique solution up to conformal coordinate transformations. It is
very convenient to choose the following form of the metric
\be
\label{lmetric}
ds^2 = -\frac{2l^2q^3 dx^+ dx^- }{(x^{-}-x^{+})^2} \, ,
\ee
which, in turn, is a way to fix the conformal coordinates
$x^{\pm}$, up to M\"obius transformations. In these coordinates
only the $t_{\pm}$ terms survive in the quantum part of the
constraints (\ref{eq4}), i.e. the semiclassical stress tensor is
just \be \label{sets} \left< T_{\pm\pm}\right> =
-\frac{N\hbar}{12\pi}t_{\pm}\> , \ee and the relevant information
of the solutions is therefore encoded in the field $\tilde{\f}$.
The crucial point  is then to choose the suitable functions
$t_{\pm}(x^{\pm})$.\\

If we want to give a description of the evaporation process for an
infalling observer very close to the horizon, the natural boundary
conditions are \cite{F,F2}
\bea
\label{cc} t_+(x^+) &=&
\frac{1}{2} \{ v,x^+ \} \, , \\ \label{mc} t_-(x^-) &=& 0 \, ,
\eea
which correspond, see eq. (\ref{sets}),
to a negative influx of radiation crossing the apparent horizon
and no outgoing flux (this is indeed what one gets in the full
four dimensional picture in fixed background considering the limit
close to the horizon). Alternatively, one can provide a
description of the evaporation process from the point of view of
an outside observer valid at late times. In this region it is
perfectly legitimate to choose the boundary conditions
\bea
\label{md}
t_+(x^+) &=& 0 \, , \\ \label{dd} t_-(x^-) &=&
-\frac{1}{2}\{ u,x^- \} \, ,
\eea
giving a positive outflux of radiation and vanishing incoming flux.
In eqs. (\ref{cc}) and
(\ref{dd})  $u$ and $v$ are to be identified with the ingoing and
outgoing Eddington-Finkelstein coordinates associated with the
dynamical Reissner-Nordstr\"om metric. Note that we cannot impose
simultaneously (\ref{mc}), (\ref{md}), otherwise the only solution
is the classical one. Therefore (\ref{cc}), (\ref{mc}) and
(\ref{md}), (\ref{dd}) are, in a sense, complementary.\\

It is worth to remark that the relations $x^+=x^+(v,\hbar)$,
$x^-=x^-(u,\hbar)$ cannot be given a priori and can only be
determined once we solve the equations. At extremality $x^+=v$ and
$x^-=u$, up to M\"obius transformations, and we have $t_v=0=t_u$.
The conditions (\ref{cc}), (\ref{dd}) imply the following form of
the ingoing and outgoing quantum fluxes
\bea
\label{flux}
\langle T^f_{vv} \rangle &=& \phantom{-} \frac{N\hbar}{24 \pi} \{ x^+, v
\} \, , \\ \langle T^f_{uu} \rangle &=& -\frac{N\hbar}{24 \pi} \{
x^-, u \} \, .
\eea
In the presence of collapsing matter and
neglecting the backreaction the ingoing flux (\ref{flux}) vanishes
for an outside observer and the outgoing flux gives the standard
Hawking radiation
\be
\label{hawk}
\langle T^f_{uu} \rangle = \frac{N\hbar}{24\pi lq^3} \tilde{m} \, ,
\ee
where $\tilde{m}$ is the total mass of the collapsing matter.\\

From the point of view of the infalling observer the solution is
the following \cite{F,F2}
\be
\tilde\phi=\frac{F(x^+)}{x^- - x^+}+\frac{1}{2}F'(x^+) \, ,
\ee
where the function $F(x^+)$ satisfies the differential equation (here we
include a general incoming matter configuration)
\be
\label{nby}
F'''=\frac{N\hbar}{24\pi}\left( -\frac{F''}{F}
+\frac{1}{2}(\frac{F'}{F})^2\right) -T_{++}^f (x^+) \, ,
\ee
and serves to relate the $x^+$ and $v$ coordinates
\be
\frac{dv}{dx^+}=\frac{lq^3}{F} \, .
\ee
The metric can also be given in the ingoing Vaidya-type gauge
\be
\label{vaie}
ds^2=-(\frac{2\tilde x^2}{l^2q^3}-l\tilde{m}(v))dv^2+2dvd\tilde x \, ,
\ee
where $\tilde x=l\tilde \phi$ and $\tilde{m}(v)$ is the deviation of the mass
from extremality. The evaporating mass function satisfies the differential
equation
\be
\label{mfv}
\partial_v \tilde{m}(v)=-\frac{N\hbar}{24\pi lq^3}\tilde{m}(v) +T_{vv}^f(v)
\, .
\ee
The negative incoming quantum flux is given by
\be
\label{prc}
\left< T^f_{vv} \right> =-\frac{N\hbar}{24\pi lq^3}\tilde{m}(v) \, .
\ee
If the incoming classical matter is turned off at some advanced time $v_f$
then the evaporating solution approaches asymptotically the extremal
configuration (up to exponentially small corrections) \cite{F,F2}
\be
\label{remnant}
\tilde{\f} = \frac{F''(x^+_{{\mathrm int}})}{2} \frac{(x^+-x_{{\mathrm
int}}^-)(x^--x_{{\mathrm int}}^-)}{x^--x^+} \, ,
\ee
where ($x^{\pm}_{{\mathrm int}}$) represent the end-point coordinates that
belong to the AdS$_2$ boundary ($x^+_{{\mathrm int}}=x^-_{{\mathrm int}}$).
For $v>v_f$ the evaporating mass follows the exponential law
\be
\label{an}
\tilde{m}(v) = \tilde{m}(v_f)e^{-\frac{N\hbar}{24\pi lq^3}(v-v_f)} \, ,
\ee
and therefore it exactly vanishes in the limit $v\to \infty$ (i.e. $x^+\to
x^+_{{\mathrm int}}$).\\

In the alternative description of the evaporation process, suitable for
the outside observer at $v=+\infty$, the solution for $\tilde \phi$ is
\be
\label{vv}
\tilde\phi= \frac{G(x^-)}{x^+ - x^-} +\frac{1}{2}G'(x^-) \, ,
\ee
where the function $G(x^-)$ verifies the differential equation
\be
\label{nbz}
G'''=-\frac{N\hbar}{24\pi}\left(
-\frac{G''}{G}+\frac{1}{2}(\frac{G'}{G})^2 \right) \, .
\ee
The most delicate point in finding a solution to the above
differential equation is the choice of the correct boundary
conditions. They come from the requirement that the two
descriptions match at the end-point $(x^+_{int},x^-_{int})$, which
belongs to both the near horizon and asymptotic regions considered
(see the conformal diagram of Fig. 1). It is worth noting,
however, that once we move away from it the corrections to eq.
(\ref{remnant}) will of course be different in the two cases and
therefore this is in agreement with the principle of
complementarity. Moreover, such a requirement is certainly
nonlocal because it implies that the form of the function $G(x^-)$
for $x^-<x^-_{int}$ (and therefore $\left< T_{uu}^f\right>$ for
finite $u$) depends on the precise form of the solution at the
end-point (where $\left< T_{uu}^f\right>=0$). Imposing that
(\ref{remnant}) be exactly
(\ref{vv}) for $x^-\to x^-_{{\mathrm int}}$ we obtain
\bea
\label{czk}
G(x^-_{{\mathrm int}}) &=& 0= F(x^+_{{\mathrm int}})
\, , \nonumber \\ \label{ccd} G'(x^-_{{\mathrm int}})&=& 0=
F'(x^+_{{\mathrm int}}) \, , 
\eea
\begin{figure}
\centerline{\psfig{figure=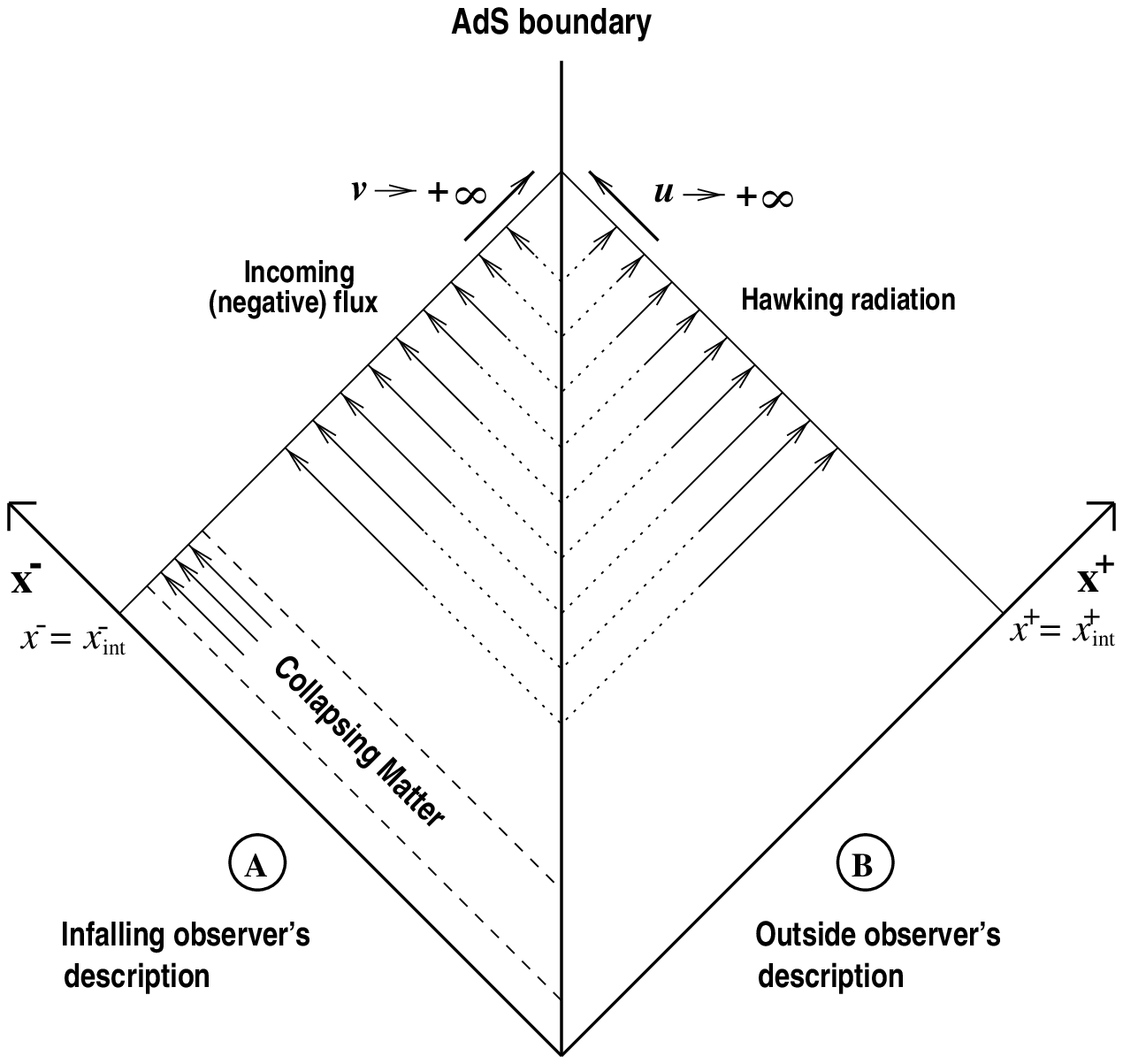,width=3.in,angle=-90}}
\end{figure}
\begin{center}
\makebox[8.5cm]{\parbox{8.5cm}{\small \noindent FIG.1. Black hole
evaporation and the Complementa\-rity Principle. Region A is the
description of the evaporation process given by an infalling
observer. Region B is the one given by an outside observer. Both
descriptions are, in a sense, complementary but agree at the
end-point $x^+_{\mathrm int}=x^-_{\mathrm int}$ ($v \rightarrow
+\infty, u \rightarrow + \infty$)}}
\end{center}
and the absence of fluxes at the end-point gives
\be
\label{cdu}
G''(x^-_{int})=-F''(x^+_{int})<0 \, .
\ee
The relation between the $x^-$ and $u$ coordinates is now given by
\be
\frac{du}{dx^-}=-\frac{lq^3}{G(x^-)} \, ,
\ee
where the minus sign
is required in order to have a positive derivative. So $F$ and $G$
are solutions of the differential equations (\ref{nby}) and
(\ref{nbz}), which in the region where $T_{vv}^f=0$ differ just
for an overall sign in their r.h.s. Moreover, both solutions have
similar boundary conditions, again up to a sign, in
$F''(x^+_{{\mathrm int}})= -G''(x^-_{{\mathrm int}})$ where
$x^+_{{\mathrm int}}=x^-_{{\mathrm int}}$. Therefore $G(x^-)$ is
functionally equal to $-F(x^+)$ after exchanging $x^+$ with $x^-$.
$F''(x^+_{{\mathrm int}})$ uniquely fixes $\tilde{m}(v_f)$ and so
(\ref{cdu}) implies that the (positive) Hawking flux is
\be
\label{cicco}
\left< T_{uu}^f(u)\right> =\frac{N\hbar}{24\pi lq^3}\tilde{m}(u)=
\frac{N\hbar}{24\pi lq^3}\tilde{m}(v_f)e^{-\frac{N\hbar}{24\pi lq^3}(u-v_f)}
\, ,
\ee
where the explicit expression for $\tilde{m}(v_f)$ is given by the formal
solution to the equation (\ref{mfv})
\bea
\label{mtil}
\tilde{m}(v_f) &=& \sum_{n=0}^{\infty}
(-\frac{N\hbar}{24\pi lq^3})^n\int_{-\infty}^{v_f}dv_1\int_{-\infty}^{v_1}dv_2
\nonumber \\
& & ....\int_{-\infty}^{v_n}dv_{n+1} T_{vv}^f(v_{n+1}) \, .
\eea
Similarly to (\ref{vaie}) the solution can now be expressed in the
outgoing Vaidya-type form
\be
ds^2=-\left( \frac{2\tilde x^2}{l^2q^3} - l\tilde{m}(u) \right)
du^2 - 2dud\tilde x \, .
\ee
It is important to point out the fact
that $\tilde{m}(v_f)$ depends on the details of the collapsing
matter through all the higher-order moments of the classical
stress tensor. We observe that for $\hbar\to 0$ $\tilde{m}(v_f)$
is the total classical mass of the collapsing matter and
(\ref{cicco}) recovers the constant thermal value of a static
near-extremal black hole (\ref{hawk}). So when backreaction
effects are neglected we loose the information of the initial
state. \\

We wish to stress that eq. (\ref{cicco}) is the first exact
calculation of the Hawking radiation flux for RN black holes at
late times which takes into account consistently backreaction
effects to all orders in $\hbar$. Our result is highly nontrivial
because two different expansions in $\hbar$ are implicit in
(\ref{cicco}), one being associated to the exponential
$e^{-\frac{N\hbar}{24\pi lq^3}(u-v_f)}$ and the other inside
$\tilde m(v_f)$, see (\ref{mtil}). While the first expansion is of
no surprise because it is nothing but the application of Stefan's
law to this particular situation, the second, consequence of our
(natural) choice of boundary conditions for the differential
equation (\ref{nbz}), is completely unexpected on physical
grounds. Actually this suggests, contrary to the earlier
predictions based on calculations made in a fixed classical
background, that the outgoing radiation may contain all the
information about the initial state already at the one-loop
semiclassical level (i.e. without recourse to a full quantization
of the theory which is still lacking). Obviously our results alone
are not enough to prove this conjecture because just from the parameter
$\tilde{m}(v_f)$ one cannot reconstruct the whole function
$T_{vv}^f$ (this is the limitation of our model which, as we have
already remarked, can give the exact Hawking flux in the RN
spacetime only at late times). Given our present achievements,
however, it is our belief that the exact $\left< T_{uu}\right>$
for all $u$ will prove to be an extremely interesting quantity.
Unfortunately the semiclassical version of spherically reduced
Einstein-Maxwell theory (without the near-horizon approximation
considered in this paper) is not solvable, so such a calculation
is a much harder challenge (recently, the full backreaction
problem for an evaporating RN black hole has been addressed numerically in 
\cite{sopi}).
Nevertheless we now know that this
quantity has to be subjected to the `boundary condition' that for
late times $u\to +\infty$ it has to reduce to eq. (\ref{cicco}).
Therefore the results obtained here cannot be confined to the
domain of validity of the particular 2d model considered, but
represent the motivation and the starting point for a full four
dimensional calculation to be performed in the physical world (we
are currently working in this direction). What is striking is
that such an input comes from the simple 2d model (\ref{paction})
and the boundary conditions (\ref{ccd}) and (\ref{cdu}).\\

It is also interesting to remark that the radiation measured by an infalling
observer and that of an external one at late times ($u\to +\infty$) are the
same under the interchange of $v$ with $u$ (which means a reflection by the
curve $x^+ = x^-$). The curve $x^+=x^-$, which is nothing but the AdS$_2$
boundary, then acts as a sort of stretched horizon \cite{STU} in the
Reissner-Nordstr\"om spacetime at least at late times when $v\to +\infty$,
$u\to +\infty$. Since the affine distance (as measured
along null rays) between the horizon and the boundary is finite, we think
that such a surface can be exactly located in the physical spacetime using
null rays. There also are indications that the degrees of freedom relevant to
account for the Bekenstein-Hawking entropy can be located at the AdS$_2$
boundary \cite{NN}. This also reinforces the idea that it could have a physical
meaning in the Reissner-Nordstr\"om spacetime.\\

This research has been partially supported by the CICYT and DGICYT, Spain.
D. J. Navarro acknowledges the Ministerio de Educaci\'on y Cultura for a FPI
fellowship. A.F. thanks R. Balbinot for useful discussions. J. N-S.
thanks the Department of Physics of Bologna University for hospitality
during the late stages of this work.

\vspace{0.5cm}
\noindent $^{\ast}$Email address: fabbria@bo.infn.it\\
\noindent $^{\dagger}$Email address: dnavarro@ific.uv.es\\
\noindent $^{\ddagger}$Email address: jnavarro@ific.uv.es
%%%%%%%%%%%%%%%%%%%%%%%%%%%%%%%%%%%%%%%%%%%%%%%%%%%%%%%%%%%%%%%%%%%%%%%%%%%%

\end{multicols}
\end{document}